\documentclass[pra,twocolumn,superscriptaddress,showpacs]{revtex4}
\usepackage{amsmath,amssymb,amsfonts,bbm,graphicx}
\newcommand{\ket}[1]{\vert #1 \rangle}

\newcommand{\braket}[2]{\langle #1 \vert #2 \rangle}

\newcommand{\ketbra}[2]{\vert #1 \rangle \! \langle #2 \vert}

\newcommand{\average}[1]{\left \langle #1  \right\rangle}
\newcommand{\be}{\begin{equation}}
\newcommand{\ee}{\end{equation}}
\newcommand{\bae}{\begin{eqnarray}} \newcommand{\eae}{\end{eqnarray}}

\def\({\left(} \def\){\right)} \def\[{\left[} \def\]{\right]}
  
\def\Tr{\hbox{Tr}}

\def\elu{\epsilon_{\hbox{\tiny U}}}
\def\eln{\epsilon_{\hbox{\tiny N}}}

\def\E{\epsilon}
\begin{document}
\title{Optimal estimation of entanglement}
\author{Marco G. Genoni}
\affiliation{Dipartimento di Fisica, Universit\`a di Milano, I-20133 Milano, Italia}
\affiliation{CNISM, UdR Milano, I-20133 Milano, Italia}
\author{Paolo Giorda}
\affiliation{I.S.I. Foundation, I-10133 Torino, Italia}
\author{Matteo G. A. Paris}
\affiliation{Dipartimento di Fisica, Universit\`a di Milano, I-20133 Milano, Italia}
\affiliation{CNISM, UdR Milano, I-20133 Milano, Italia}
\affiliation{I.S.I. Foundation, I-10133 Torino, Italia}
\date{\today}
\begin{abstract}
Entanglement does not correspond to any observable and its evaluation
always corresponds to an estimation procedure where the amount of
entanglement is inferred from the measurements of one or more proper
observables.  Here we address optimal estimation of entanglement in the
framework of local quantum estimation theory and derive the optimal
observable in terms of the symmetric logarithmic derivative. We evaluate
the quantum Fisher information and, in turn, the ultimate bound to
precision for several families of bipartite states, either for qubits or
continuous variable systems, and for different measures of entanglement.
We found that for discrete variables, entanglement may be efficiently
estimated when it is large, whereas the estimation of weakly entangled
states is an inherently inefficient procedure. For continuous variable
Gaussian systems the effectiveness of entanglement estimation strongly
depends on the chosen entanglement measure.  Our analysis makes an
important point of principle and may be relevant in the design of
quantum information protocols based on the entanglement content of
quantum states.  
\end{abstract}
\maketitle
\section{Introduction}\label{s:intro}
Entanglement is perhaps the most distinctive feature of quantum
mechanics, and definitely the most relevant resource for quantum
information processing \cite{EntR}. Indeed, quantification of
entanglement and schemes for its measurement have been the subject of
extensive efforts in the last decade
\cite{Aud06,Che03,Eis07,Guh04,Guh07,
Sun07,Lou06,Wal06,Ren04,Nav08,Hor03,Aci00,Hor02,Dur01,Bar03,Guh03,Dar03,
Pla02,Pit03}. Indeed, the entanglement content of a quantum state is a
crucial piece of information in the design of quantum information
protocols, and a question naturally arises on whether quantum mechanics
itself poses limits to the precision of its determination.  As a matter
of fact, any quantitative measure of entanglement corresponds to a
nonlinear function of the density operator and thus cannot be associated
to a quantum observable.  As a consequence, any procedure aimed to
evaluate the amount of entanglement of a quantum state is ultimately a
parameter estimation problem, where the value of entanglement is
indirectly inferred from the measurement of one or more proper
observables. An optimization problem thus naturally arises, which may be
properly addressed in the framework of quantum estimation theory (QET)
\cite{QET}, which provides analytical tools to find the optimal
measurement according to some given criterion. 
\par
Our aim is indeed to evaluate the ultimate bounds to precision posed by
quantum mechanics, {\em i.e} the smallest value of the
entanglement that can be discriminated, and to determine the optimal
measurements achieving those bounds. Being entanglement an intrinsic
property of quantum states we adopt local quantum estimation theory,
where the optimal estimators are those maximizing the Fisher information
\cite{Hel67,BC94,BC96} and in turn minimizing the variance at fixed
value of entanglement.  Local QET provides any family of quantum states
with a geometric structure based on distinguishability and accounts for
the optimal measurement that can be performed on the quantum system as
well as the optimal data processing of the outcomes of the measurement.
\par
Local QET has been applied to the estimation of quantum optical phase
\cite{Mon06} as well as to estimation problems involving non unitary processes
in open quantum systems \cite{Sar06}, either in finite dimensional
systems \cite{Hot06} or continuous variable ones \cite{Mon07}. This
includes the optimal estimation of the noise parameter of depolarizing
\cite{Fuj01} or amplitude-damping \cite{Zhe06,Mon07} channels.
Recently, the geometric structure induced by the Fisher
information itself has been exploited to give a quantitative operational
interpretation for multipartite entanglement \cite{Boi08} and to assess
quantum criticality as a resource for quantum estimation \cite{ZP07}. 
\par
In this paper we systematically apply local QET to the problem of
efficiently estimate the amount of entanglement of a quantum state.  We
consider several families of bipartite states, either for qubits or
Gaussian states, and evaluate the symmetric logarithmic derivative to
estimate entanglement through different measures, {\em e.g.} negativity
or linear entropy.  Then we explicitly calculate the quantum Fisher
information and derive the ultimate bounds to the precision of
estimation. Overall, we found that, both for qubits and Gaussian states,
entanglement may be efficiently estimated when it is large. On the other
hand the estimation of a small amount of entanglement for qubits is an
inherently inefficient procedure, {\em i.e} the signal-to-noise ratio is
vanishing for vanishing entanglement, whereas for Gaussian states it
depends on the chosen measure of entanglement.  We also found that the
presence of other free parameters besides entanglement does not
generally influence the estimation precision, thus preventing the
possibility of further optimizing the estimation procedure.
\par
The paper is structured as follows: in the next Section we give some
basic elements of quantum estimation theory and introduce the quantum 
signal-to-noise to assess the estimability of a parameter. In Section
\ref{s:qubits} we analyze the
estimation of entanglement by means of negativity \cite{NVW}
and linear entropy for the family of pure two-qubit states as well as
for two families of entangled mixtures. In Section \ref{s:PPT} we
address a family of PPT bound-entangled states \cite{HorodPPT} for
two-qutrit systems as an example of states with an inherently small
amount of entanglement.  In Section \ref{s:CV} we address entanglement
estimation for Gaussian states, either pure states (twin-beams) or
entangled mixtures.  Section \ref{s:conclusions} closes the paper with
some concluding remarks. 
\section{Quantum estimation theory}\label{s:qet}
In an estimation problem one tries to infer the value the of a parameter
$\lambda$ by measuring a different quantity $X$, which is somehow
related to $\lambda$. An estimator $\hat\lambda \equiv\lambda 
(x_1, x_2,..)$ for $\lambda$ is a real function of the outcomes of 
measurement. The Cramer-Rao theorem
\cite{Cra46} establishes a lower bound for the variance
${\mathrm{Var}}(\lambda)$ of any unbiased estimator 
\begin{align}
{\mathrm{Var}}(\lambda) \geq \frac{1}{M F(\lambda)} \label{eq:CramerRao}
\end{align}
in terms of the number of measurements $M$ and the so-called 
Fisher Information (FI) 
\begin{align}
F(\lambda) &=\sum_x  p(x\vert \lambda)\, [\partial_\lambda \ln p(x\vert
\lambda)]^2 
= \sum_x \frac{[\partial_\lambda  p(x\vert
\lambda)]^2}{p(x\vert \lambda)}  , \label{eq:ClassicalFisher}
\end{align}
where $p(x\vert\lambda)$ denotes the conditional probability of
obtaining the value $x$ when the parameter has the value $\lambda$.
\par
In quantum mechanics, according to the Born rule we have
$p(x\vert \lambda) = \Tr[E_x \varrho_\lambda]$ where $\left\{E_x\right\}$
are the elements of a positive operator-valued measure (POVM) and
$\varrho_\lambda$ is the density operator parametrized by the quantity
we want to estimate. Introducing the Symmetric Logarithmic
Derivative (SLD) $L_\lambda$ as the operator satisfying
the equation \begin{align}
\frac{L_\lambda \varrho_\lambda + \varrho_\lambda L_\lambda}{2} =
\frac{\partial \varrho_\lambda}{\partial \lambda} \label{eq:SLD}
\end{align}
we have that
$\partial_\lambda p(x\vert\lambda) = \Tr[ \partial_\lambda\varrho_\lambda E_x]
= \hbox{Re}( \Tr[\varrho_\lambda L_\lambda E_x])$, and 
the Fisher Information in Eq. (\ref{eq:ClassicalFisher}) may 
be rewritten as
\begin{align}
F(\lambda) = \sum_x \frac{\hbox{Re}(\Tr[\varrho_\lambda L_\lambda E_x])^2}
{\Tr[\varrho_\lambda E_x]} .\label{eq:CQFisher} 
\end{align}
Starting from Eq. (\ref{eq:CQFisher}) one may prove that $F(\lambda)$
is upper bounded by the so-called \emph{Quantum Fisher Information}
(QFI) \cite{BC94, BC96} 
\begin{align}
F(\lambda) \leq H(\lambda) \equiv \Tr[\varrho_\lambda L_\lambda^2]\:,
\label{eq:QuantumFisher} \end{align}
and, in turn, that $\hbox{Var} (\lambda) \geq [M H(\lambda)]^{-1}$
represents the quantum version of the Cramer-Rao theorem, {\em i.e.}
the ultimate bound to precision for any quantum measurement aimed 
to estimate the parameter $\lambda$. The SLD itself provides an optimal 
measurement, that is, using a measurement described by the projectors 
over the eigenbasis of $L_\lambda$ we saturate inequality (\ref{eq:QuantumFisher}). 
\par
Upon diagonalizing $\varrho_{\lambda} = \sum_k p_k 
|\psi_k\rangle\langle\psi_k|$ and using  Eqs. (\ref{eq:SLD}) 
and (\ref{eq:QuantumFisher}) we obtain
\begin{align}\label{defs}
L_{\lambda} & = 2 \sum_{n,m}
\frac{\langle\psi_n | \partial_\lambda \varrho_\lambda|\psi_m\rangle }
{p_n + p_m} |\psi_n\rangle\langle\psi_m| \\ 
H (\lambda ) & = \sum_n \frac{(\partial_\lambda p_n)^2}{p_n} + 2\sum_{n,m}
\frac{(p_n-p_m)^2}{p_n+p_m} |\langle\psi_n|\partial_\lambda\psi_m\rangle|^2
\nonumber\, 
\end{align}
which, for a family of pure states $\varrho_\lambda=
|\psi_\lambda\rangle\langle\psi_\lambda|$ reduces to 
\begin{align}\label{defsp}
L_\lambda & =2\,\partial_\lambda |\psi_\lambda\rangle\langle\psi_\lambda|
\\ H(\lambda) & = 4 \left[
\langle\partial_\lambda\psi_\lambda
|\partial_\lambda\psi_\lambda\rangle
+ \langle\partial_\lambda\psi_\lambda |
\psi_\lambda\rangle^2
\right]
\nonumber\,.
\end{align}
When more than a parameter are involved we have quantum states
$\varrho_{\Lambda}$ depending on a set of $N$ parameters $\Lambda=\{
\lambda_j \}$, $j=1,\dots,N$. In this case the geometry of the
estimation problem is contained in the QFI matrix, whose elements are
defined as ${\boldsymbol H}(\Lambda)_{ij} =
\frac{1}{2}\Tr[\varrho_{\Lambda} \{ L_i,L_j \}]$ where $L_i$ is the SLD
that corresponds to the parameter $\lambda_i$ and $\{A,B\}=AB + BA$
denotes anti-commutator. The explicit formula for the QI matrix reads as
follows
\begin{align}
&H(\Lambda)_{ij} = \sum_n \frac{(\partial_i p_n) (\partial_j p_n)}{p_n} + 
\sum_{n,m}
\frac{(p_n-p_m)^2}{p_n + p_m} \times \nonumber \\
&\times (\langle \psi_n|\partial_i \psi_m\rangle
\langle \partial_j \psi_m| \psi_n \rangle +
\langle \psi_n|\partial_j \psi_m\rangle
\langle \partial_i \psi_m| \psi_n \rangle )
\end{align}
The inverse of the Fisher matrix provides a lower bound ${\boldsymbol \gamma} 
\geq H^{-1}$, on the covariance matrix ${\boldsymbol \gamma}_{ij}= 
\langle \lambda_i \lambda_j \rangle - \langle \lambda_i\rangle\langle
\lambda_j\rangle$ of global estimators of $\Lambda$, which is not 
generally achievable. On the other
hand, the diagonal elements of the inverse Fisher matrix provide
achievable bounds for the variances of single parameter estimators
(at fixed value of the others)
\begin{align}
{\mathrm{Var}}(\lambda_i) = {\boldsymbol \gamma}_{ii} 
\geq ({\boldsymbol H}^{-1})_{ii}.
\label{eq:QCRMulti}
\end{align}
Let us now suppose to reparametrize the family of quantum states with
a new set of parameters $\tilde{\Lambda}= \{\tilde{\lambda}_j=
\tilde{\lambda}_j(\Lambda) \}$. We have
$\tilde{\partial}_j = \sum_i B_{ij} \partial_i$ with $B_{ij} =
\partial \lambda_i/\partial \tilde{\lambda}_j$ and, in turn  
\begin{align}
\tilde{L}_j & = \sum_i B_{ij} L_i \nonumber \\ 
H(\tilde{\Lambda})_{ij} & = \sum_{r,s}
B_{ir}
H(\Lambda)_{rs} 
B_{js} \label{eq:Hchangevar}
\end{align}
{\em i.e.} $\widetilde{{\boldsymbol H}} = {\boldsymbol B} 
{\boldsymbol H} {\boldsymbol B}^T$. \par
In the following we address the problem of finding the bounds to the
estimation of the entanglement between the two subsystems $A,B$ of a
family of bipartite quantum states. The general strategy will be to
start from the expression of the family of states in terms of a given
set of "natural" parameters and then make a change of variable in order
to write the state directly in terms of the chosen entanglement
monotone $\lambda=\epsilon$.
Once this is achieved, the results obtained hold, at every fixed value of
the entanglement,  for the whole orbit of states that can be obtained
by acting on the given one with local unitary operators $U_{AB}\doteq U_A\otimes U_B$.
Indeed, the latter, acting locally on the two subsystems $A$ and $B$,
they do not change the entanglement of the state; furthermore they do
not change the value of the QFI.  This follows from the observation
that, if a given $L_{\epsilon}$ is the solution of (\ref{eq:SLD}), the
SLD that will correspond to $\tilde{\varrho}_{\epsilon}=
U_{AB}\varrho_{\epsilon}U_{AB}^\dagger$ is given by
$\tilde{L}_{\epsilon}=U_{AB}L_{E}U_{AB}^\dagger$ and, due to the cyclic property
of the trace,
$H(\epsilon)=\Tr[\varrho_{\epsilon}L_{\epsilon}^2]=
\Tr[\tilde{\varrho}_{\epsilon}\tilde{L}_{\epsilon}^2]$.
\par
We finally notice that in order to assess the estimability of a given
parameter the relevant figure of merit is given by the signal-to-noise
ratio
\begin{align}
R(\lambda) &= \frac{\lambda^2}{\mathrm{Var}(\lambda)} \le
Q (\lambda)  = \lambda^2 H(\lambda) \label{eq:SNR def}\;,
\end{align}
rather than the variance itself. In particular the SNR of an estimator
is relevant to assess its performances in estimating small values of the
parameter. Eq. (\ref{eq:SNR def}) shows that
the SNR $R(\lambda)$ is bounded by the quantum signal-to-noise ratio
(QSNR) $Q(\lambda)$ expressed in terms of the QFI.
Upon taking into account repeated measurements we have that 
the number of measurements leading to a $99.9 \%$ ($3\sigma$) 
confidence interval corresponds to a relative
error 
$$
\delta^2 = \frac{9 \hbox{Var}(\lambda)}{M\lambda^2} = \frac{9}{M}
\frac1Q(\lambda) = \frac{9}{M\lambda^2 H(\lambda)}
$$
Therefore, the number of measurements needed to achieve a $99.9\%$
confidence interval with a relative error $\delta$ scales as
\begin{equation}
M_\delta(\lambda) = \frac9{\delta^2} \frac1{Q(\lambda)} \label{eq:Mdelta}
\end{equation}
In other words, a vanishing $Q(\lambda)$ implies a diverging number 
of measurements to achieve a given relative error, whereas for a 
finite  $Q(\lambda)$ the number of measurements is determined by the
desired level of precision. 
In order to have a non-vanishing $Q(\lambda)$ for small value
of the parameter $\lambda$, the QFI should diverge at least as
$H(\lambda) \sim \lambda^{-2}$ for vanishing $\lambda$.
We notice that a similar quantity, namely $\lambda H(\lambda)$,
has been used in to
asses estimation strategy for the parameter of a qubit depolarizing
channel \cite{Fuj03}.
\section{Two-qubit systems} \label{s:qubits}
In this section we analyze estimation of entanglement for families of
two-qubit states and evaluate limits to precision using the formalism
developed in the previous section. At first we address the set of pure
states and then consider families of entangled mixtures. In both cases 
we consider different measures of entanglement.
\subsection{Pure states}
We start by considering the set of pure states of two qubits. 
Upon exploiting the Schmidt decomposition
\be
\ket{\Psi_q}=\sqrt{q}\ket{0}_A\ket{0}_B+\sqrt{1-q}
\ket{1}_A\ket{1}_B\:, \label{eq: Psi_q}
\ee
the whole family of pure states can be parametrized by a single parameter:
the Schmidt coefficient $q$. Since for two-qubit pure states $q$ is
itself an entanglement monotone, all measures of entanglement can be
expressed as a monotone function $\epsilon=\epsilon(q)$. As a
consequence, in order to determine the precision of estimation it 
suffices to evaluate the QFI $H(q)$ and then use the rule for
repametrization in Eq. (\ref{eq:Hchangevar}), {\em i.e.}
$H(\epsilon)=H[q(\epsilon)][\partial_\epsilon q(\epsilon)]^2$.
Since the states are pure, the SLD may be evaluated as $L_{q}=2\partial_q
\ketbra{\Psi_q}{\Psi_q}$; the resulting Cramer-Rao bound and the QSNR 
read as follows 
\begin{align}
{\mathrm{Var}}(q)& \geq H(q)^{-1}=q(1-q) \\ 
Q(q) &= \frac{q}{1-q} \stackrel{q\rightarrow 0}{\sim} q \:. 
\label{eq:QSNR twoqubitpure}
\end{align}
$Q(q)$ vanishes for vanishing $q$, thus indicating that any
estimator of the Schmidt coefficient $q$ becomes
less and less precise for vanishing $q$. It is worth noting that
the Schmidt coefficient $q$ coincides with the only independent
eigenvalue of the reduced density matrix $\varrho_{A(B)}=\hbox{diag}\{q,1-q\}$,
which is diagonal in the Schmidt basis. Therefore the QSNR in Eq.
(\ref{eq:QSNR twoqubitpure}) also imposes bound to the determination
of the eigenvalue of $\varrho_{A(B)}$. Indeed, the same bound could have
been obtained by applying the estimation machinery directly
to $\varrho_{A(B)}$.
\par
Let us now consider two different measures of entanglement for 
pure two-qubit states, {\em i.e} the negativity $\epsilon_{\hbox{\tiny N}}$
\cite{NVW} and the (normalized) linear entropy
$\epsilon_{\hbox{\tiny L}}=2(1-\hbox{Tr}[\varrho_A^2])$. In terms
of the Schmidt coefficient $q$ we have
\be
\epsilon_{\hbox{\tiny N}}=
\sqrt{\epsilon_{\hbox{\tiny L}}}=2\sqrt{q(1-q)}
\ee
We recall that the negativity is a good measure of entanglement for
generic two-qubit states, {\em i.e} it is an entanglement monotone
and it differs from zero iff the state is entangled, whereas 
the linear entropy is a good entanglement monotone only iff the state is pure.
Upon expressing the Schmidt coefficient as 
$q=\frac12 (1-\sqrt{1-\epsilon_{\hbox{\tiny N}}^2})$ 
and using (\ref{eq:Hchangevar}), we have
$\mathrm{Var}(\epsilon_{\hbox{\tiny N}}) \geq
H(\epsilon_{\hbox{\tiny N}})^{-1}=
1-\epsilon_{\hbox{\tiny N}}^2$, 
$\mathrm{Var}(\epsilon_{\hbox{\tiny N}}) \geq 
H(\epsilon_{\hbox{\tiny L}})=
4\epsilon_{\hbox{\tiny L}}(
1-\epsilon_{\hbox{\tiny L}})$ and, in turn, 
\begin{align}
Q(\epsilon_{\hbox{\tiny N}}) & =\frac{\epsilon_{\hbox{\tiny
N}}^2}{1-\epsilon_{\hbox{\tiny N}}^2}\stackrel{\epsilon_{\hbox{\tiny
N}}\rightarrow 0}{\sim} \epsilon_{\hbox{\tiny N}}^2 \\
Q(\epsilon_{\hbox{\tiny L}}) & =\frac{\epsilon_{\hbox{\tiny
L}}}{4(1-\epsilon_{\hbox{\tiny L}})} \stackrel{\epsilon_{\hbox{\tiny
L}}\rightarrow 0}{\sim} \epsilon_{\hbox{\tiny L}}/4\:.
\end{align}
The optimal estimator for the Schmidt coefficient has a variance
${\mathrm{Var}}(q)$ which is minimum for $q=0,1$ (product state) and
maximum for $q=1/2$ (Bell state) whereas for the two entanglement
measure we have that ${\mathrm{Var}}(\epsilon_{\hbox{\tiny N}})$ is
monotonically decreasing with $\epsilon_{\hbox{\tiny N}}$;
${\mathrm{Var}}(\epsilon_{\hbox{\tiny L}})$ is minimum when the state is
either in a product form ($\epsilon_{\hbox{\tiny L}}=0$) or is maximally
entangled ($\epsilon_{\hbox{\tiny L}}=1$) and is  maximum in the
"intermediate" case ($\epsilon_{\hbox{\tiny L}}=1/2$).  Despite the
variances behave quite differently we have the same qualitative
behavior of the quantum signal-to-noise ratios and of the number of
measurement necessary at fixed relative error $M_\delta$. Indeed, in all
cases the QSNR is an increasing function of the parameter and it
diverges when the latter takes its maximum value ($q=1$,
$\epsilon_{\hbox{\tiny N}}=\epsilon_{\hbox{\tiny L}}=1$);
$M_\delta(\epsilon)$ diverges for $\epsilon=\epsilon_{\hbox{\tiny N}}
=\epsilon_{\hbox{\tiny L}}=0$ and then decreases monotonically, going to
zero for the maximum value of entanglement
$\epsilon=\epsilon_{\hbox{\tiny N}}=\epsilon_{\hbox{\tiny L}}=1$ .
Moreover for vanishing entanglement the QNSR of the linear entropy
estimator is vanishing slower than the corresponding quantity for the
negativity.  We conclude that the linear entropy is a more efficient
entanglement estimator though, being the QSNR vanishing, the estimation
is anyway inherently inefficient.
\par
The above result can obviously generalized to the case of systems
composed by a qubit and an $N$-level system; indeed only two of the
dimensions of the latter can be used to express the state: the reduced
density matrices of both subsystems have only two non zero eigenvalues.
\subsection{Entangled mixtures}
We now consider few families of mixed entangled states with
different properties and show that they exhibit a common
behavior concerning estimation of entanglement.
The first family is described by the set of density matrices
\be
\varrho=U(q) \nu_p U^\dag (q)\label{eq: Mixedstate1}
\ee
where 
\begin{align}
\nu_p & =p\ketbra{0,0}{0,0}+(1-p)\ketbra{1,1}{1,1} \nonumber \\ 
U(q) & =\exp\{i\,\hbox{arcos}\, q\, \sigma_x\otimes \sigma_x\} \nonumber \,.
\end{align}
These states depend on two parameters $(q,p)$ and
are obtained via the action of the entangling operator
$U(q)$ on the classically correlated state $\nu_p$. Upon varying the
parameter $p$ we may control the purity $\mu(p)=1-2p(1-p)$
of the state, while varying
$q$ we tune the amount of entanglement. The QFI matrix is
diagonal with elements:
\bae
{\boldsymbol H}(p,q) =
\hbox{diag}\left(\frac{1}{p(1-p)},\frac{(1-2p)^2}{q(1-q)}\right)
\eae
The negativity of the state  (\ref{eq: Mixedstate1}) is given by
the unique
negative eigenvalue of the partially transposed state $\varrho^{T_A}$:
\be
\epsilon_{\hbox{\tiny N}}=2\sqrt{q(1-q)} (1-2p)
\ee
Upon inverting the above relation and the one for the purity, 
we reparametrize the set of
states in terms of the new parameter $(p,q) \rightarrow 
(\mu,\epsilon_{\hbox{\tiny N}})$.
The transfer matrix is given by
\begin{equation}
{\boldsymbol B}=\left(\begin{array}{cc}
-\frac{1}{2\sqrt{2\mu -1}} & \frac{\epsilon_{\hbox{\tiny N}}^2}
{2\sqrt{(2\mu -1)^2(2\mu -1 -\epsilon_{\hbox{\tiny N}}^2)}}\\
0 & \frac{\epsilon_{\hbox{\tiny N}}}
{2\sqrt{(2\mu -1)(2\mu -1 -\epsilon_{\hbox{\tiny N}}^2)}}
\end{array}\right)
\end{equation}
and the inverse QFI matrix:
\begin{equation}
{\boldsymbol H}(\mu,\epsilon_{\hbox{\tiny N}})^{-1}=\left(\begin{array}{cc}
-4\mu^2 + 6\mu -2 
&2\epsilon_{\hbox{\tiny N}}(1-\mu)\\
2\epsilon_{\hbox{\tiny N}}(1-\mu) & 1-\epsilon_{\hbox{\tiny N}}^2
\end{array}\right)
\end{equation}
The corresponding bound on the variance is thus given by:
\be
{\mathrm{Var}}(\epsilon_{\hbox{\tiny N}})\ge
\left({\boldsymbol H}(\mu,\epsilon_{\hbox{\tiny N}})^{-1}\right)_{22} =
1-\epsilon_{\hbox{\tiny N}}^2\,
\label{varn}
\ee
which represents the bound to the precision of any entanglement
(negativity) estimation procedure performed at fixed purity $\mu$.
The result in Eq. (\ref{varn}) is independent on the purity, 
no optimization procedure may be pursued, and it coincides with
the bound obtained and discussed in the previous
subsection for pure states.
\par
\begin{figure}[h!]
\includegraphics[width=0.46\textwidth]{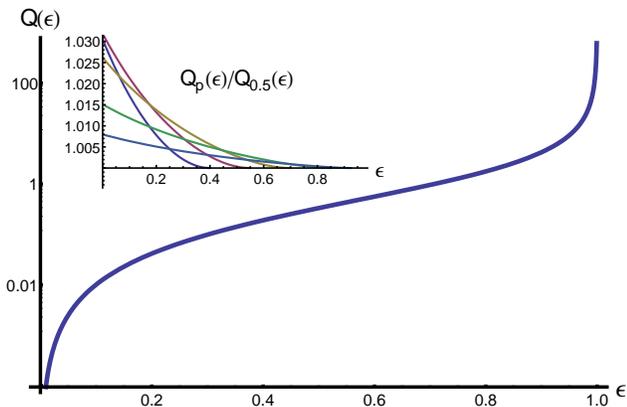}
\caption{Quantum signal-to-noise ratio for the estimation of
entanglement (negativity) of a two-qubit Werner state as a
function of the negativity at fixed $q=0.5$. The inset shows 
the ratio $Q_{p}(\epsilon_{\hbox{\tiny N}})/
Q_{q=0.5}(\epsilon_{\hbox{\tiny N}})$ for different value 
of $p$.}
\label{fig:SNR_Wer_Neg}
\end{figure}

Let us now consider the family of Werner-like states
\be
\varrho_{pq} = \frac{1-p}{4}\openone \otimes \openone + p \ketbra{\Psi_q}{\Psi_q}
\ee
obtained by depolarizing an entangled state $\ket{\Psi_q}$ 
of the form given in Eq. (\ref{eq: Psi_q}).  This set of states 
depends on two parameters $p,q$. As in the previous example upon varying
the parameter $p$ we may control the purity $\mu(p)=(1+3p^2)/4$ of the state,
while the amount of entanglement depends on both parameters.
The eigenvalues of $\varrho_{pq}$ depends only on $p$ whereas
the eigenvectors depends only on $q$. The QFI matrix is thus given 
by the diagonal form
\be
{\boldsymbol H}(p,q)=\mbox{diag}\left \{ \frac{3}{1+(2-3p)p},\, \,
\frac{p^2}{q(1-q)(1+p)} \right \}\,
\ee
and the inverses of the diagonal elements correspond
to the ultimate bounds to ${\mathrm{Var}}(p)$ and ${\mathrm{Var}}(q)$
of any estimator of $p$ and $q$, either at fixed value of the other
parameter or in a joint estimation procedure. Entanglement of Werner 
states may be evaluated in terms of negativity,
\be
\epsilon_{\hbox{\tiny N}}=\max \left\{0,
\frac12 \left[p\left((1+4\sqrt{q(q-1)}\right)-1\right]\right\}
\label{wen}
\:,
\ee
which implies that Werner states are entangled for $(1 > p > [1 
+ 4q (1 - q)]^{-1}$.
Upon inverting Eq. (\ref{wen}) for $p$ or $q$ we may parametrize 
the Werner states using $(\epsilon_{\hbox{\tiny N}},q)$ or 
$(p,\epsilon_{\hbox{\tiny N}})$ and evaluate the QFI matrices 
$H(\epsilon_{\hbox{\tiny N}},q)$ and $H(p,\epsilon_{\hbox{\tiny N}})$, 
their inverses and, in turn, the corresponding bounds to the precision
of entanglement (negativity) estimation. 
The main results are that the ultimate bounds to the variance, 
and thus to the QSNR depend very slightly on the other free
parameter ($q$ or $p$). In other words, estimation procedures
performed at fixed value of $p$ or $q$ respectively shows 
different precision, but the differences are negligible in the whole
range of variations of the parameters.
We do not report here the analytic expression of $Q(\epsilon_{\hbox{\tiny N}})$
at fixed $p$ or $q$ which is quite cumbersome. Rather, we show the behavior 
of $Q(\epsilon_{\hbox{\tiny N}})$ in Fig. \ref{fig:SNR_Wer_Neg}.
On the left we show the QSNR $Q_{q=0.5}(\epsilon_{\hbox{\tiny N}}) 
$ for $q=0.5$, whereas on the right we show the ratio 
$Q_{p}(\epsilon_{\hbox{\tiny N}})/Q_{q=0.5}(\epsilon_{\hbox{\tiny N}})$
for different value of $p$ (a similar behavior may be observed 
upon varying $q$).
As it is apparent from the main panel the QSNR is a growing function 
of $\epsilon_{\hbox{\tiny N}}$, vanishes for
vanishing negativity and diverges for maximally entangled states
$\epsilon_{\hbox{\tiny N}}=1$.  The inset shows that 
there is almost no
dependence on the actual value of $p$ and $q$ respectively and
this prevents any possible optimization of the estimation 
procedure. For small $\eln$ we have 
$Q (\eln) \simeq f(q) \eln^2$ and $Q(\eln) \simeq g(p)\eln^2$
respectively, where both the functions $f(q)\simeq 1$ and 
$g(p)\simeq 1$ are again very close to unit value for the whole ranges of
variation of $q$ and $p$. 
\par
In all the cases we have considered, the QSNR is small for the most
part of entanglement range and starts growing only for highly 
entangled states. In other words, estimation of
entanglement is, on average, an inefficient procedure.
\section{Two-qutrit bound entangled states} \label{s:PPT}
\begin{figure}[h!]
\includegraphics[width=0.22\textwidth]{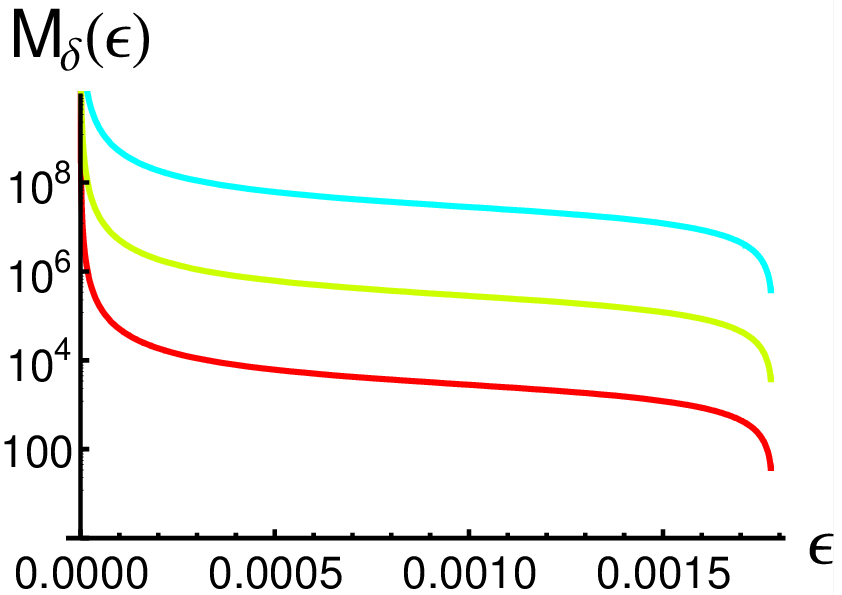}
\includegraphics[width=0.22\textwidth]{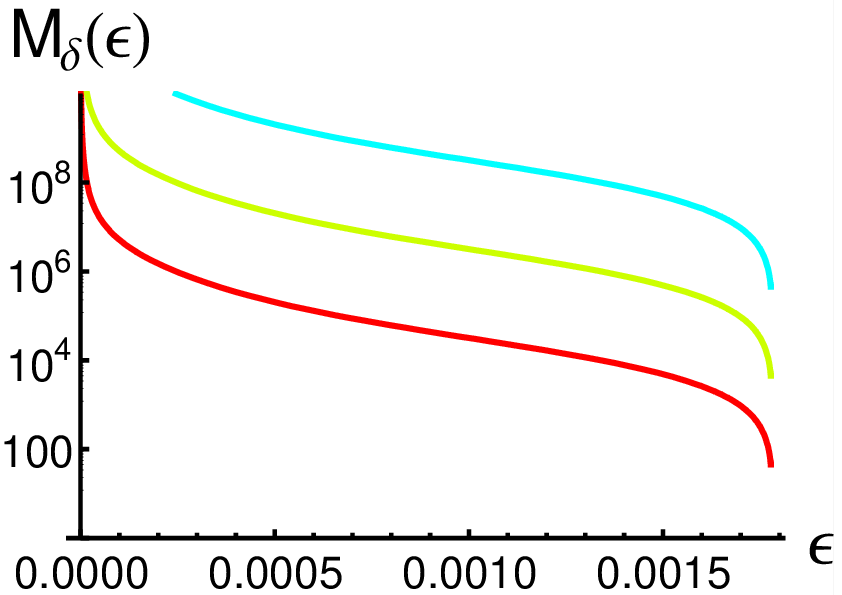}
\caption{Number of measurement $M_\delta(\elu)$ needed to achieve a
given relative error $\delta=10^{-1}$ (red), $\delta=10^{-2}$ (green), 
$\delta=10^{-3}$ (blue) in the the estimation of LUR entanglement
measure $\elu$ of PPT state $\varrho_a$ in Eq.(\ref{eq:bound}).
Left: $a \in [0,4/13]$, Right: $a \in [4/13,1]$}\label{fig:M_CLUR_PPT}
\end{figure}

In the previous section we have seen how the estimation of entanglement,
as measured by negativity is a fairly inefficient procedure for weakly
entangled states. Here we want to test how the QFI and the related
bounds behave when one considers states that have inherently small
amount of entanglement. A paradigmatic examples of such states are the
so called {\em bound entangled states}, which exhibit non-classical 
correlations even if they satisfy the separability criterion based 
on partial transposition of the density matrix 
\cite{HorodPPT,PeresSep,HorodSep}. The first example of bound entangled
states is given by the following family of two spin-1 states 
\cite{HorodPPT} 
\begin{eqnarray}
\label{eq:bound}
\varrho_a&=& \frac{a}{1+ 8 a} \big(
   | \downarrow 0 \rangle\langle \downarrow 0  | +
   |\downarrow \uparrow \rangle\langle \downarrow  \uparrow | + 
   |0 \downarrow \rangle\langle  0 \downarrow | 
   \nonumber \\ &+&
   |0 \uparrow \rangle\langle  0 \uparrow | +
   |\uparrow 0 \rangle\langle \uparrow 0 | \big)
  + \frac{3 a}{1+ 8 a}
   |E \rangle\langle E | 
\nonumber \\ &+&
    \frac{1}{1+ 8 a}
   |\Pi \rangle\langle \Pi |
 \end{eqnarray}
where
\begin{align}
| E \rangle & =
\frac{1}{\sqrt{3}}\left(
| \downarrow\downarrow
\rangle + | 00 \rangle + |\uparrow\uparrow\rangle
\right) 
\nonumber \\ 
|\Pi \rangle & =
\sqrt{\frac{1+a}{2}} | \uparrow\downarrow \rangle +
\sqrt{\frac{1-a}{2}} |\uparrow\uparrow \rangle.
\nonumber  
\end{align}
Since for all values of the parameter $a$ $\; \varrho_a$ has a
positive partial transpose (PPT), negativity  cannot be used as a
measure of the quantum correlations present in state. In order to
estimate the entanglement we will use the scheme proposed in
\cite{HoffmanLUR,HoffmanLURPPT}. The latter is based on the following
considerations. Given the sets of $n$ non-commuting operators $\{A_i\}$
and $\{B_i\}$ acting locally on the subsystem A and B respectively one
has a lower bound for the sum of the local uncertainties relations
(LUR): 
\be
\sum_i \delta A_i^2 > U_A \mbox{ and } \sum_i \delta B_i^2 > U_B
\ee
where $\delta O_i^2 = \average{O_i^2}-\average{O_i}^2$ is the variance
of the operator $O_i$. Since for all separable states on has that
$\sum_i \delta (A_i +B_i)^2> U_A+U_B$, the latter inequality set a
necessary condition for a state to be entangled. The relative violation
of the inequality defined as 
\be
\elu=1-\frac{\sum_i \delta (A_i +B_i)^2}{U_A+U_B}
\ee
can then be used as a measure of the quantum correlations present in the
given state. The violation is necessary condition for the presence of
the entanglement, thus, in order to effectively have and maximize such
violation one can judiciously choose and optimize the choice of the sets
$\{A_i\}$ and $\{B_i\}$. The result of a possible optimization for the
state $\varrho_a$ is given in \cite{HoffmanLURPPT} and the
corresponding relative violation depends on the parameter $a$ and is
given by: 
\be
\elu(\varrho_a)=\frac{3 a^2(1-a)}{4 (2+a)(1+8a)^2}
\ee
The latter expression can be used to parameterize the state
$\varrho_a$ in terms of $\elu(\varrho_a)$ and then
apply the QFI machinery in order to obtain the desired bound on the
estimation of relative violation of the LUR. The results 
are shown in Fig. \ref{fig:M_CLUR_PPT}.
We first note that the relative violation of the LUR is small, {\em i.e} 
$\elu(\varrho_a)\in[0,2/1125]$. Nonetheless, the number of
measurements $M_\delta(\elu)$ is of the same order of those needed in the 
qubit case $M_\delta(\epsilon_{\hbox{\tiny N}})$ in similar conditions,
and thus the overall efficiency of the estimation process is comparable
for most part of the entanglement range. 
Finally, we notice that also for this family of qutrits the number 
of measurements $M_\delta(\elu)$ diverges for vanishing entanglement.
\section{Gaussian states}\label{s:CV}
In this section analyze continuous variable systems and derive the
bounds for the estimation of entanglement of two-mode Gaussian states 
\cite{GaussiansAOP}.
After a brief introduction  we consider both pure states, {\em i.e.}
twin-beam and the family of mixed states represented by squeezed thermal
states (STS). Different measures of entanglement will be considered.
\par 
The characteristic function of the state $\varrho$ is defined as 
$\chi[\varrho]({\bf \Gamma})= \hbox{Tr}\left[\varrho\, 
D(\Gamma)\right]$ where $D(\Gamma)=\exp{ \left [i \textbf{R}^T 
{\bf \Omega} {\bf \Gamma} \right ] }$ is the displacement
operator, defined in terms of the 
symplectic matrix
\be
{\bf \Omega}=\bigoplus_{k=1}^n \left( \begin{array}{cc}
0 & 1 \\
-1 & 0 \\
\end{array}\right )_n
\ee
and the vector $\textbf{R}^T=(q_1,p_1,\cdots,q_n,p_n)$ of canonical 
operators. ${\bf \Gamma}^T=(a_1,b_1,\cdots,a_n,b_n)$ denotes the Cartesian 
coordinates and we have $\left [ R_k,R_h\right ]=i {\bf \Omega}_{kh}$. 
Two-mode Gaussian state are those with a characteristic function of the
form
\be
\chi[\varrho]({\bf \Gamma})=\exp{\left( -\frac{1}{2}{\bf \Gamma}^T{\bf \sigma}{\bf \Gamma}
+i{\bf \Gamma}^T{\bf \overline{X }}\right)}
\ee
where
\be
{\bf \sigma}_{kh}=\frac{1}{2}\langle\{R_k,R_h\}\rangle-\langle R_k\rangle\langle R_h\rangle
\ee
is the covariance matrix of the second moments and ${\bf \overline{X }}$
denotes mean values. The second term in the exponential does not contain 
any information about entanglement, and can be set to zero via local
operation.  The covariance matrix completely characterize the
state and by means of local symplectic transformations 
can be transformed into the standard block form :
${\sigma}=  \left(\begin{array}{cc}
A & C \\ C & B 
\end{array}\right )$ 
where $A=\hbox{Diag}(a,a)$, $B=\hbox{Diag}(b,b)$, and
$C=\hbox{Diag}(c_+,c_-)$.
For two-mode Gaussian states PPT condition is necessary and sufficient
for separability and thus the entanglement properties of the state are 
encoded in the least symplectic eigenvalue \cite{GaussianCrit}
\be
\tilde{d}_-=\sqrt{(a-c_+)(a+c_-)}
\ee
where the symplectic spectrum of ${\bf \sigma }^{T_A}$ can be evaluated by finding the
eigenvalues of ${\bf \Omega^{-1}} {\bf \sigma}^{T_A} .$ In this framework the PPT
criterion can be cast in term of the smallest symplectic eigenvalue i.e.,
$\varrho$ is separable iff $\tilde{d}_- \ge 1/2$.
Indeed, $\tilde{d}_-$ is itself an entanglement monotone. Furthermore,
all the different entanglement measures for symmetric Gaussian states
that have been proposed \cite{NVW, EoFGiedke,EntBuresMarian1}
turns out to be a monotone function of the smallest symplectic
eigenvalue. Since we are interested in the estimation of entanglement,
we may first study the estimation of the symplectic eigenvalue $\tilde{d}_{-}$
and then use repametrization in order to asses the performances of other 
entanglement monotones. In particular, we will focus on the following
measures:
\begin{align}
\epsilon_{\hbox{\tiny N}}(\tilde{d}_-) &= \max{\{0, -\ln{2}\tilde{d}_-\}} 
\label{eq:LogNeg} \\
\epsilon_{\hbox{\tiny L}}(\tilde{d}_-) &= 1 - \frac{4\tilde{d}_-}{1+4\tilde{d}_-^2} 
\label{eq:LinEnt} \\
\epsilon_{\hbox{\tiny S}}(\tilde{d}_{-}) &= 1 -2 \tilde{d}  
\label{eq:BuresEnt} \\
\epsilon_{\hbox{\tiny B}}(\tilde{d}_{-}) &= \frac{(1-\sqrt{2 \tilde{d}})^2}
{1+2\tilde{d}}. 
\label{eq:SEnt}
\end{align}
where the expressions for the
linear entropy $\epsilon_{\hbox{\tiny L}}$ has been obtained
in the pure state case. In particular $\epsilon_{\hbox{\tiny L}}$
and the logarithmic negativity $\epsilon_{\hbox{\tiny N}}$ will be 
used for pure 
states, whereas the Bures distance-based measure 
$\epsilon_{\hbox{\tiny B}}$ and $\epsilon_{\hbox{\tiny S}}$ 
\cite{EntBuresMarian1} 
will be used for mixed states.
Notice that $\epsilon_{\hbox{\tiny B}}$ is a good measure of entanglement
just for symmetric two-mode Gaussian states and that Eq.  (\ref{eq:BuresEnt}) 
has been obtained for this particular class of states.
In order to evaluate the QFI, one has first to determine the actual
expression of $\tilde{d}_{-}$. Then, one expresses the elements of the
covariance matrix in terms of the chosen entanglement monotone and
proceeds with the repametrization rules described in Section
\ref{s:qet}. 
\subsection{Pure states}
Here we address estimation of the entanglement of pure two
modes Gaussian states, {\em i.e.} twin-beam. These are defined by the 
following relations between the elements of the covariance matrix: 
$a=b$ i.e., the states are symmetric and $c_+=-c_-=\sqrt{a^2-1/4}$. 
The states can thus be described by the single parameter $a$ or, by 
inverting $\tilde{d}_-= a -\sqrt{a^2-1/4}$ they can be completely 
described by their entanglement content. For Gaussian states the 
evaluation of the SLD and the QFI, besides the use of Eqs. (\ref{defs}), 
may be pursued using phase-space techniques.
In fact, for pure states we have
$L_\E=2\partial_\E \varrho_\E$ and this allows to directly evaluate 
the characteristic function of the SLD as follows 
\be
\chi[L_\E] (\bf\Gamma )=2\partial_\E\chi[\varrho_\E]=
-{\bf \Gamma}^T \partial_\E\sigma_\E{\bf \Gamma}\ \chi[\varrho_\E]\:.
\ee
where ${\bf \Gamma}^T=({\bf \Gamma_1}^T, {\bf \Gamma_2}^T)$.
The corresponding QFI
$H(\epsilon) = \mbox{Tr}[\varrho\, L_\E^2]$ is given by
\bae
H(\epsilon)&=&
\int\!\!\int
\frac{d^{2}{\bf \Gamma}_1 }{2\pi}\frac{d^{2}
{\bf \Gamma}_2}{2\pi}\, \mathcal{I}
\left(\E,{\bf \Gamma}_1,{\bf \Gamma}_2\right)
\eae 
where the integrand function reads as follows
\bae
\mathcal{I}&=& \chi[L_\E]({\bf \Gamma}_1)\chi[L_\E]({\bf \Gamma}_2)
\mbox{ Tr }[\varrho D^\dagger({\bf \Gamma}_1) D^\dagger({\bf \Gamma}_2)]
\label{eq: IGauss1}
\eae
We now use the relations $D[{\bf \Gamma}]^\dagger=D[-{\bf \Gamma}]$ 
and $D({\bf \Gamma}_1) D({\bf \Gamma}_2)=D({\bf \Gamma}_1+
{\bf \Gamma}_2)g({\bf \Gamma}_1,{\bf \Gamma}_2)$ and we 
rewrite (\ref{eq: IGauss1}) as
\bae
\mathcal{I}&=&
\left({\bf \Gamma}_1^T \partial_\E\sigma_\E{\bf \Gamma}_1\right)
\left({\bf \Gamma}_2^T\partial_\E\sigma_\E{\bf \Gamma}_2\right)
\cdot\nonumber \\ &&
\chi_{\varrho_\E}({\bf \Gamma}_1)
 \chi_{\varrho_\E}
({\bf \Gamma}_2) \chi_{\varrho_\E:}(-{\bf \Gamma}_1-{\bf \Gamma}_2)
g({\bf \Gamma}_1,{\bf \Gamma}_2)
\label{eq: IGauss2}
\nonumber \\ &=&
\left ({\bf \Gamma}^T{\bf \Sigma_1}{\bf \Gamma}\right )
\ \left ( {\bf \Gamma}^T{\bf \Sigma_2}{\bf \Gamma} \right )
\ \exp{\left (-\frac{1}{2}{\bf \Gamma}^T{\bf \Delta}{\bf \Gamma} \right )}
\eae
where we have introduced  the matrices:
\be
{\bf \Sigma}=  \left( \begin{array}{cc}
2\sigma &\sigma \\
\sigma & 2\sigma
\end{array}\right ),
{\bf \Sigma_1}=  \left( \begin{array}{cc}
\partial_\E\sigma &0 \\
0 & 0
\end{array}\right ),
{\bf \Sigma_2}=  \left( \begin{array}{cc}
0 &0 \\
0 & \partial_\E \sigma
\end{array}\right )\nonumber
\ee
and where $\Delta= {\bf \Sigma} + i \Upsilon$, with $\Upsilon=\frac12\,
\sigma_y \otimes {\mathbbm I} \otimes \sigma_y$ in terms of Pauli matrices.  
The result of the
integration QFI is a function of $\E,\partial_\E \sigma_\E$ and can be
expressed in various ways depending on the entanglement monotone that
one chooses to estimate.  By setting
$a=\frac{1}{2}\cosh\epsilon_{\hbox{\tiny N}}$ in the covariance matrix,
one can use the logarithmic negativity $\epsilon_{\hbox{\tiny N}}$. In
this case one finds that the QFI is independent on the entanglement
content of the state.  \par If one uses directly the symplectic
eigenvalue $\tilde{d}_-$ the QFI now depends on the entanglement
monotone:
\be
H(\tilde{d}_-)=\tilde{d}_-^{-2}
\label{Gnu_}
\ee 
and the minimal variance in the estimation can be obtained in the limit
of infinite entanglement i.e., $\tilde{d}_-=0$. Moreover we observe
that, while for the logarithmic negativity one sees that the QSNR
$Q(\eln)$ is simply proportional to $\eln^2$, if we consider the least
symplectic eigenvalue we have $Q(\tilde{d}_-)=1$ over the whole range of
variation, \emph{i.e.} the estimation procedure can be done efficiently
either for highly entangled states and weakly entangled ones.
\par
As a matter of fact, twin-beam may be also written in the Fock basis as 
$\ket{\Psi} = \sum_n f_n \ket{n}\ket{n}$
where, in terms of the log-negativity or the linear entropy 
one may write
\begin{align}
f_n  = \sqrt{2 \frac{(\cosh \epsilon_{\hbox{\tiny
N}}-1)^n}{(\cosh \epsilon_{\hbox{\tiny N}}+1)^{1+n)}}}\: 
= \sqrt{2\frac{
\epsilon_{\hbox{\tiny L}}^n
(1-\epsilon_{\hbox{\tiny L}})
}{
(2-\epsilon_{\hbox{\tiny L}})^{1+n}
}}\,.
\end{align}
Using this representation we may directly exploit Eqs. (\ref{defsp}):
for a generic parameter $x$ and $f_n(x)\in \mathbb{R}$
we have $\braket{\Psi}{\partial_x \Psi}=0$ and thus
\begin{equation}
H(x)=\braket{\partial_x \Psi}{\partial_x \Psi}=\sum_n (\partial_x
f_n(x))^2\:.
\end{equation}
Using the above equation one recover the result for the 
log-negativity and may evaluate the QFI in terms of the
linear entropy obtaining
\begin{equation}
H(\epsilon_{\hbox{\tiny L}}) =
\left (4(2 -\epsilon_{\hbox{\tiny L}})(\epsilon_{\hbox{\tiny L}}-1)^2 
\epsilon_{\hbox{\tiny L}}  \right )^{-1} .
\end{equation}
\subsection{Entangled mixtures}
We now analyze entanglement estimation for a relevant 
family of mixed Gaussian states labeled by two independent 
parameters. The symmetric two-mode squeezed thermal states 
(STS) are given by
\begin{align}
\varrho_{ST} = S_{12}(r,\phi) (\nu_{N_t}\otimes\nu_{N_t}) S_{12}^{\dag}(r,\phi)
\label{eq:ST}
\end{align}
and represents a two parameter family 
$\varrho_{ST}=\varrho_{ST}(N_t,r)$ obtained from symmetric 
two-mode thermal state with $N_{t}$
thermal photons for each mode by the action of the two-mode 
squeezing operator
$S_{12}(r) = \exp \{ r (a_1^{\dag}a_2^{\dag} - a_1a_2)\}$. 
The family in Eq. (\ref{eq:ST}) occurs when one considers
the propagation of twin-beam in a noisy channel or the 
generation of entanglement from a noisy background \cite{ntw}, 
and represents the CV generalization of the family of entangled 
mixed states introduced in Eq. (\ref{eq: Mixedstate1}).
We evaluate the corresponding Fisher information matrix 
${\boldsymbol H}(r,N_t)$ and obtain a diagonal matrix
\begin{align}
{\boldsymbol H}(r,N_t) =
\hbox{diag}\left( 8 - \frac{4}{1+ 2 N_t(1+N_t)},\frac{2}{N_t(1+N_t)}\right) .
\end{align}
Let us now consider the smallest symplectic eigenvalue 
$\tilde{d}_{-}(r,N_t)$ and the purity of the state $\mu(N_t)$
\begin{align}
\tilde{d}_{-} = \frac{e^{-2r}}{2} (1 + 2 N_t) \qquad
\mu = \frac{1}{2 N_t +1}. \nonumber
\end{align}
Upon inverting the above equations  
we may reparametrize the set of states in terms of
the new parameters $(\tilde{d}_{-},\mu)$, the transfer matrix 
${\boldsymbol B}$ being given by  
\begin{align}
{\boldsymbol B} &=
\left(
\begin{array}{c c}
-\frac{1}{2 \tilde{d}_{-}} & 0 \\
-\frac{1}{2 \mu} & -\frac{1-\mu}{2\mu^2} - \frac{1}{2\mu}
\end{array}
\right) \nonumber
\end{align}
The new QFI matrix ${\boldsymbol H}(\tilde{d}_{-},\mu)$
is calculated by means of Eq. (\ref{eq:Hchangevar})
and the bound on the covariance matrix 
${\boldsymbol \gamma}(\tilde{d}_{-},\mu) \geq {\boldsymbol
H}(\tilde{d}_{-},\mu)^{-1}$ is established by its inverse
\begin{align}
{\boldsymbol H}(\tilde{d}_{-},\mu)^{-1} &=
\left(
\begin{array}{c c}
\tilde{d}_{-}^2 & -\frac{\tilde{d}_{-}\mu(1-\mu^2)}{2} \\
-\frac{\tilde{d}_{-}\mu(1-\mu^2)}{2} & \frac{\mu^2}{2}(1-\mu^2)
\end{array}
\right)
\end{align}
The lower bounds on the variance for the symplectic eigenvalue
is given by
\begin{align}
{\mathrm{Var}}(\tilde{d}_{-}) &\geq ({\boldsymbol H}
(\tilde{d}_{-},\mu)^{-1})_{11} = \tilde{d}_{-}^2
\end{align}
and represents the limit to the precision of any estimator of $\tilde{d}_{-}$
at fixed purity $\mu$. In particular, we observe that this bound does not 
depend on the purity, and coincides with the bound  in Eq. (\ref{Gnu_})
obtained for pure states. Therefore also for this class of states
the QSNR is $Q(\tilde{d}_{-})=1$ and hence $\tilde{d}_-$ can be always 
estimated efficiently. 
\par
Let us now consider a generic measure of entanglement 
$\epsilon=\epsilon(\tilde{d}_{-})$. Upon using Eq. (\ref{eq:Hchangevar})
we may show that the reparametrization $(\tilde{d}_{-},\mu)\rightarrow
(\epsilon, \mu)$ leads to
\begin{align}
H(\epsilon,\mu)_{11}
& = \left( \frac{\partial \tilde{d}}{\partial \epsilon}
\right)^2 H(\tilde{d}_{-},\mu)_{11} \label{eq:FishFunct} \\
{\mathrm{Var}}(\epsilon) &\geq
({\boldsymbol H}(\epsilon,\mu)^{-1})_{11}
=
\left( \frac{\partial \tilde{d}}{\partial \epsilon} \right)^{-2}
({\boldsymbol H}(\tilde{d}_{-},\mu)^{-1})_{11} \label{eq:CRBFunct}
\end{align}
Let us consider the two monotone functions of the symplectic eigenvalue
$\epsilon_{\hbox{\tiny S}}(\tilde{d}_-)$
and $\epsilon_{\hbox{\tiny B}}(\tilde{d}_-)$ introduced in Eqs. (\ref{eq:SEnt}) 
and (\ref{eq:BuresEnt}).
The symplectic eigenvalue can be expressed in terms of the measures as
\begin{align}
\tilde{d}(\epsilon_{\hbox{\tiny S}}) &= \frac{1 - \epsilon_{\hbox{\tiny S}}}{2}
\nonumber \\
\tilde{d}(\epsilon_{\hbox{\tiny B}}) &= \frac{1 + 2 \epsilon_{\hbox{\tiny B}} -
 \epsilon_{\hbox{\tiny B}}^2 -2\sqrt{2 \epsilon_{\hbox{\tiny B}} - 
 \epsilon_{\hbox{\tiny B}}^2}}{2(1-\epsilon_{\hbox{\tiny B}})^2}.
\nonumber
\end{align}
thus leading to
\begin{align}
{\mathrm{Var}}(\epsilon_{\hbox{\tiny S}}) &\geq (1-\epsilon_{\hbox{\tiny S}})^2
\nonumber \\
{\mathrm{Var}}(\epsilon_{\hbox{\tiny B}}) &\geq \frac{\epsilon_{\hbox{\tiny B}}
(2-\epsilon_{\hbox{\tiny B}})(1-\epsilon_{\hbox{\tiny B}})^2}{4} \nonumber
\end{align}
We notice that ${\mathrm{Var}}(\epsilon_{\hbox{\tiny S}})$ 
and ${\mathrm{Var}}(\epsilon_{\hbox{\tiny B}})$ 
show different behavior; in particular, while the bound
on ${\mathrm{Var}}(\epsilon_{\hbox{\tiny S}})$ vanishes 
only when $\epsilon_{\hbox{\tiny S}}$
is maximum ($\epsilon_{\hbox{\tiny S}}=1$), the bound on
${\mathrm{Var}}(\epsilon_{\hbox{\tiny B}})$ reaches zero both when
$\epsilon_{\hbox{\tiny B}}$ is maximum ($\epsilon_{\hbox{\tiny B}}=1$) and when
is minimum ($\epsilon_{\hbox{\tiny B}}=0$) and presents
 a maximum for $\epsilon_{\hbox{\tiny B}}=1-1/\sqrt{2}$.
\par
We finally evaluate the QSNR for the measures of entanglement
introduced, obtaining
\begin{align}
Q(\epsilon_{\hbox{\tiny S}}) &\leq \frac{\epsilon_{\hbox{\tiny S}}^2}
{(1-\epsilon_{\hbox{\tiny S}})^2} \stackrel{\epsilon_{\hbox{\tiny
S}}\rightarrow 0}{\sim} \epsilon_{\hbox{\tiny S}}^2  \\
Q(\epsilon_{\hbox{\tiny B}}) &\leq \frac{4 \epsilon_{\hbox{\tiny B}}}
{(1-\epsilon_{\hbox{\tiny B}})^2(2-\epsilon_{\hbox{\tiny B}})}
\stackrel{\epsilon_{\hbox{\tiny
B}}\rightarrow 0}{\sim} 2\epsilon_{\hbox{\tiny B}}  \nonumber
\end{align}
The two QSNRs are increasing function of entanglement, vanish 
for zero entanglement and diverge for maximally entangled states. 
In turn, the numbers of measurements  $M_\delta(\epsilon_{\hbox{
\tiny B}})$ and  $M_\delta(\epsilon_{\hbox{\tiny S}})$ vanish for 
maximum entanglement and diverge for vanishing entanglement. 
The QSNR of $\epsilon_{\hbox{\tiny B}}$ is vanishing slower than the 
corresponding quantity for $\epsilon_{\hbox{\tiny S}}$
and therefore we conclude that the measure based on the
Bures distance is more efficiently estimable 
compared to the linear measure $\epsilon_{\hbox{\tiny S}}$.
On the other hand, being 
the QSNR vanishing, the estimation is
anyway inherently inefficient. 
\section{Conclusions} \label{s:conclusions}
Entanglement of quantum states is not an observable quantity. 
On the other hand, the amount of entanglement can be indirectly 
inferred by an estimation procedure, {\em i.e.} by measuring 
some proper observable and then processing the outcomes by a 
suitable estimators. In this paper we have established a first 
approach to the estimation of the entanglement content of a 
quantum state and to the search of optimal quantum estimators, 
{\em i.e} those with minimum variance. Our approach is based 
on the theory of local quantum estimation and allows, upon the 
evaluation of the quantum Fisher information, to derive the 
ultimate bounds to precision imposed by quantum mechanics. 
We have applied our analysis to several families of quantum 
states either describing finite size systems or continuous 
variables ones, and have considered different measures in order
to quantify the amount of entanglement. \par
For the case two-qubit pure state we have found that 
any procedure to estimate entanglement (either quantified by 
negativity or by linear entropy) is efficient only for maximally
or near maximally entangled states, whereas it becomes inherently 
inefficient for weakly entangled states. In particular, the number 
of measurements needed to achieve a $99.9\%$ confidence interval 
withing a given relative error diverges as far as the value of 
entanglement becomes small. The same results hold also for 
families of mixed states, remarkably for the orbit of an entangling
unitary an for a general class of Werner-like states. 
Indeed in all the examples we have considered the presence of 
other free parameters besides entanglement, though changing the 
QFI, does not affect the estimation precision, \emph{i.e.} the value
of the relevant element of the inverse QFI matrix. In turn, this also
prevents the possibility of further optimizing the estimation procedure.
\par
On the other hand, we have showed that for an important class of
states whose entanglement of distillation is zero (PPT bound entangled
states), the use of an optimized measure of quantum correlation i.e.,
the relative violation of local unitary relations introduced in
\cite{HoffmanLURPPT}, results in a more efficient estimation
procedure, with precision comparable with those achievable in the
estimation of entanglement through negativity. 
\par
In the case of continuous variable Gaussian states we have shown 
that the estimation of the least symplectic eigenvalue $\tilde{d}_-$ of
the covariance matrix may be performed with arbitrary precision 
at fixed number of measurements, independently on the value of $\tilde{d}_-$
itself and for both pure states and mixed states. 
If we rather introduce other measures of entanglement proposed in
literature, in particular the logarithmic negativity for pure
states and the one based on the Bures distance \cite{EntBuresMarian1}
for the symmetric squeezed thermal (mixed) states, we observe the same behavior
obtained in the discrete variable case: the estimation is efficient only
for maximally entangled state and inherently inefficient for weakly
entangled states.  Therefore it is apparent that for continuous variable
systems, the efficiency of the estimation strongly depends on the
measure one decides to adopt.
\par
In conclusion, upon exploiting the geometric theory of quantum
estimation we have quantitatively evaluated the ultimate bounds posed by
quantum mechanics to the precision of entanglement estimation for
several families of quantum states. To this aim we used the quantum
Cramer-Rao theorem and the explicit evaluation of the quantum Fisher
information matrix.  We have also given a recipe to build the
observable achieving the ultimate precision in terms of the symmetric
logarithmic derivative.  The analysis reported in this paper  makes an
important point of principle and may be relevant in the design of
quantum information protocols based on the entanglement content of
quantum states.  Finally, we notice that our approach may be generalized
and applied to the estimation of other quantities not corresponding to
proper quantum observables, as the purity of a state or the coupling 
constant of an interaction Hamiltonian \cite{ZP07,MK08}.  Work along 
this lines is in progress and results will
be reported elsewhere. 
\section*{Acknowledgments}
This work as been partially supported by the CNR-CNISM convention.

\end{document}